# A Multi-stage Error Diagnosis for APB Transaction

Automated APB Transaction Error Detection for 2025 CAD Contest at ICCAD

CHENG-YANG TSAI*

Dept. of Computer Science & Engineering, Yuan Ze University

s1111505@mail.yzu.edu.tw

TZU-WEI HUANG*

Dept. of Computer Science & Engineering, Yuan Ze University

s1113315@mail.yzu.edu.tw

JEN-WEI SHIH*

Dept. of Computer Science & Engineering, Yuan Ze University

s1113343@mail.yzu.edu.tw

I-HSIANG WANG*

Dept. of Computer Science & Engineering, Yuan Ze University

s1113335@mail.yzu.edu.tw

YU-CHENG LIN*

Dept. of Computer Science & Engineering, Yuan Ze University

linyu@saturn.yzu.edu.tw

RUNG-BIN LIN*

Dept. of Computer Science & Engineering, Yuan Ze University

csrlin@saturn.yzu.edu.tw

Functional verification and debugging are critical bottlenecks in modern System-on-Chip (SoC) design, with manual detection of Advanced Peripheral Bus (APB) transaction errors in large Value Change Dump (VCD) files being inefficient and error-prone. Addressing the 2025 ICCAD Contest Problem D, this study proposes an automated error diagnosis framework using a hierarchical Random Forest-based architecture. The multi-stage error diagnosis employs four pre-trained binary classifiers to sequentially detect Out-of-Range Access, Address Corruption, and Data Corruption errors, prioritizing high-certainty address-related faults before tackling complex data errors to enhance efficiency. Experimental results show an overall accuracy of 91.36%, with near-perfect precision and recall for address errors and robust performance for data errors. Although the final results of the ICCAD 2025 CAD Contest are yet to be announced as of the submission date, our team achieved first place in the beta stage, highlighting the method's competitive strength. This research validates the potential of hierarchical machine learning as a powerful automated tool for hardware debugging in Electronic Design Automation (EDA).

**CCS CONCEPTS** • Hardware → Verification → Functional verification • Computing methodologies → Machine learning → Machine learning approaches → Classification and regression trees • Hardware → Electronic design automation

**Additional Keywords and Phrases:** Error Diagnosis, Machine Learning, Hierarchical Classification, VCD Analysis

## 1 INTRODUCTION

Driven by Moore's Law, the continual increase in chip complexity has caused functional verification and debugging for Systems-on-Chip (SoCs) to consume more resources than the design itself, creating a "verification crisis." This directly impacts the product's time-to-market (TTM), making verification efficiency the decisive factor in a project's success or failure.

A key bottleneck in this challenge lies in post-simulation failure triage, a process that typically involves engineers performing time-consuming manual inspection of vast waveform datasets [1]. Although the Value Change Dump (VCD) format, a widely supported IEEE standard [2, 3], can capture the rich behavior of a design during simulation, the files it generates are often extremely large, sparse, and contain millions of data points. This makes manual analysis not only inefficient and error-prone but also a major bottleneck in the entire design cycle [3]. Recent advancements have highlighted the need for automated tools to address this issue. For instance, clustering-based approaches have been proposed to group related failures in regression verification, improving triage efficiency by categorizing failures for further analysis [4]. Similarly, automated failure triage frameworks that unify root-cause localization and failure binning have shown promise in reducing manual effort [5]. These developments underscore the urgent need for automated analysis tools to streamline the debugging process.

Complex classification tasks often challenge single, monolithic classifiers. A more effective strategy is a hierarchical approach that decomposes a problem into simpler, sequential sub-tasks. This allows the model to efficiently filter easy cases and focus computational resources on more difficult ones, establishing a new paradigm for EDA fault diagnosis. For instance, automated assertion generation has been developed to enhance verification by creating targeted checks, reducing the scope of manual inspection [6].

This hierarchical concept is successfully demonstrated by cascaded architectures in computer vision, such as Cascade R-CNN for object detection [7]. Its multi-stage refinement process significantly improves accuracy while mitigating overfitting. Drawing inspiration from this, our study posits that different types of bus transaction errors can be similarly filtered and identified by a sequence of classifiers. This approach aligns with modern hardware verification strategies, like reinforcement learning for coverage-driven test generation, which also prioritize tasks to optimize verification efficiency [8].

This multi-stage processing concept also aligns with the development direction of AI-native EDA. The "Large Circuit Models" (LCMs) vision proposed by Chen et al. is essentially a multi-modal, multi-stage learning framework[9]. LCMs aim to create a unified circuit representation by fusing and aligning data from different design stages, from high-level functional specifications to detailed physical layouts [9]. Each stage in the design flow, such as logic synthesis or physical design, can be considered an independent "modality" requiring specific representation learning strategies. This concept of aligning representations across different abstraction levels throughout the entire design flow provides strong theoretical support for the multi-model hierarchical diagnosis method proposed in this study. Furthermore, recent work on mining simulation metrics for failure triage has demonstrated the potential of data-driven approaches to prioritize and distribute failures effectively, reinforcing the need for multi-stage frameworks in EDA [10].

Ensuring the integrity of bus protocols like the Advanced Peripheral Bus (APB) is crucial in SoC design, as errors such as Address Corruption, Data Corruption, and Out-of-Range Access can cause severe system failures. Manually inspecting large VCD files to find these errors is a time-consuming and error-prone task, highlighting the need for automation. While techniques like unsupervised clustering have been explored to accelerate post-silicon debug [11], we propose a machine learning method based on a cascaded architecture. Our approach uses a series of specialized models to sequentially detect the three error types. This layered method improves both accuracy and efficiency by filtering erroneous transactions early, building on prior work in machine learning-based anomaly detection for post-silicon debugging [1].

In the domain of hardware verification, past research has proposed several innovative methods for post-silicon debugging. For example, novel trigger generation methods [12], and error diagnosis methods based on multiple inconsistent executions [13] have improved debugging efficiency and accuracy from the perspectives of data analysis, trigger conditions, and execution behavior, respectively. Additionally, recent advancements in clustering-based failure triage for RTL regression debugging have introduced dynamic weighting approaches to enhance failure categorization [4]. Building on this foundation, our research addresses the class of problems presented in Problem D of the ICCAD Contest 2025 [14] by incorporating a cascaded machine learning architecture to enhance automated diagnostic capabilities, offering a precise and efficient solution for modern SoC verification challenges.

## 2 PROBLEM DESCRIPTION

This paper aims to address the challenges presented in Problem D of the ICCAD Contest 2025[14], with a core focus on automated error diagnosis for the Advanced Microcontroller Bus Architecture (AMBA) in (SoC) design. Within this framework, the Advanced Peripheral Bus (APB) is designed as a low-power, low-bandwidth bus with a simple interface, primarily used for connecting the processor to various low-speed peripherals. APB employs a simple Master-Slave architecture, which we refer to in this paper as Requester and Completer. A basic APB write or read transaction involves a set of core signals, including: PADDR (address bus), PWDATA (write data bus), PRDATA (read data bus), PSEL (peripheral select signal), PENABLE (enable signal), and PWRITE (read/write control signal). The correct interpretation of the temporal behavior of these signals in VCD files is fundamental to automated error diagnosis.

Our cascaded approach designs a series of specialized models for the three error types defined in ICCAD Contest 2025 Problem D: Out-of-Range Access, Address Corruption, and Data Corruption. Since Address Corruption and Data Corruption are large-scale systemic errors that often require inspecting multiple APB transactions for detection, our proposed multi-stage error diagnosis can efficiently filter these broad-ranging errors, thereby enhancing the efficiency of subsequent error handling processes. The characteristics of these three errors will be detailed below.

Table 1: Characteristics of APB Transaction Errors

| Error Category | Affected Signal(s) | Causal Mechanism |
| --- | --- | --- |
| Out-of-Range Access | PADDR | Requester logic error |
| Address Corruption | PADDR | Short circuit between adjacent address lines |
| Data Corruption | PWDATA/ PRDATA | Short circuit between adjacent data lines |

Out-of-Range Access is defined as a protocol-level violation. It occurs when a requester initiates a transaction, but its target address PADDR falls outside the valid address space declared by the target completer. Suppose the valid memory address range for Completer 1 is designed to be from 0x00 to 0x7F. However, due to a logic error at the Requester side, it initiates an access request to this completer for address 0xC2 (11000010). Since 0xC2 is not within the valid range, the Completer cannot respond correctly. This error leads to unpredictable system behavior. For a write operation, the data will not be stored in the intended memory; for a read operation, the completer may return an indeterminate X value. In worse scenarios, it could lead to critical data being overwritten or a system crash.

Address Corruption is defined as a physical-level error where the signal integrity of the address bus PADDR is compromised. This is typically caused by hardware manufacturing defects, such as a short circuit or an open circuit between adjacent wires. A Requester writes data to address 0x88 (10001000). However, at the Completer, due to an unintended physical short between its address lines a5 and a4, these two bits cannot change independently. When the Requester drives 0x88, the short circuit effect causes the address actually received by the Completer to become 0x98 (10011000). The impact of Address Corruption is extremely subtle and destructive. Data is written to a valid but unintended address, which may not only overwrite important data at that location but also cause latent errors in subsequent system operations. The root cause of such errors is very difficult to trace because the manifestation of system failure may be far removed in time and logic from the initial erroneous write operation.

Data Corruption is defined as a physical-level error similar to Address Corruption, but it affects the write data bus PWDATA. In this case, the target address of the transaction is correct, but the data is altered during transmission. When the Requester writes data 0xD5 (11010101) to address 0x8A (10001010), Completer 1, due to a short between d3 and d2, actually receive the data as 0xDD (11011101). The result is that the intended data 0xD5 is incorrectly stored as 0xDD, leading to errors in subsequent reads. Data Corruption directly undermines the system's data integrity. Incorrect data is stored in memory or registers, and when this corrupted data is read and used in subsequent computations or control flows, it will trigger a series of functional errors.

In summary, although these three error types all lead to eventual functional failure in the VCD waveform, their root causes and behavioral characteristics have subtle yet critical differences. Therefore, this problem constitutes an ideal application scenario for machine learning. By building a general-purpose, automated diagnostic model capable of quickly and accurately classifying these well-defined yet subtly featured errors from raw VCD data, we can accelerate the critical analysis process and improve the efficiency of subsequent error handling.

## 3 METHODOLOGY

This study addresses the challenge of diagnosing APB transaction errors in the 2025 ICCAD Contest Problem D: APB Transaction Recognizer: Recognizing from VCD Waveforms. We propose a multi-stage diagnosis method that leverages four pre-trained binary classification models to progressively detect different types of errors. This hierarchical diagnostic approach is designed to enhance the accuracy and efficiency of error detection.

### 3.1 Data Sample Generation

To train and evaluate our models, we generated a large number of APB transaction test samples based on the official definitions of erroneous transactions from the 2025 ICCAD Problem D. The data samples strictly adhere to the official specifications, ensuring that the error components are fully consistent with their definitions. Each data sample consists of 20 APB transactions, with each transaction containing a 32-bit address (Xaddress) and 32-bit data (Xdata), accompanied by a label (Y) indicating its error class.

### 3.2 Model Training and Evaluation

Each pre-trained model was trained and tested using 200,000 data samples, with an 80%/20% split for the training and test sets. An independent binary classifier was constructed for each error type to ensure high-precision detection. The input and classification tasks for each model are as follows.

- out_of_range detection: Input Xaddress, classifies as out_of_range_error vs. no_error.
- address_error detection: Input Xaddress, classifies as address_error vs. no_error.
- data_error_0 detection: Input Xdata, classifies as data_error_0 vs. non_data_error_0.
- data_error_1 detection: Input Xdata, classifies as data_error_1 vs. non_data_error_1.

To enhance the detection capability for data-type errors, we introduced the following experimental terms and used them during model training (this subdivision is for experimental design, and the final results will combine the two into a single data_error category for presentation).

- data_error_0: An error caused by floating, where two bits in the data are constant across 20 transactions, and their binary values are consistently 00.
- data_error_1: An error caused by floating, where two bits in the data are constant across 20 transactions, and their binary values are consistently 11.
- non_data_error_0: The set of negative samples for training the data_error_0 model, defined as samples that are not data_error_0, including no_error and data_error_1.
- non_data_error_1: The set of negative samples for training the data_error_1 model, defined as samples that are not data_error_1, including no_error and data_error_0.

### 3.3 Choice of Machine Learning Algorithm

The Random Forest (RF) algorithm was selected as the primary model for all detection tasks. This choice was based on comparative experiments with alternative algorithms such as Support Vector Machines,

Gradient Boosting, and Logistic Regression. RF consistently achieved higher accuracy and AUC scores for this dataset. Its advantages include.

- Robustness to noise and correlated features, allowing stable performance in high-dimensional sensor data.
- Capability to capture nonlinear relationships through ensemble decision trees.
- Strong empirical results in both training and cross-validation phases.

### 3.4 Hyperparameters

To balance model performance and generalization, the following RF hyperparameters were adopted.

- n_estimators = 200: Increases model stability and reduces variance.
- max_depth = 15: Prevents overfitting while maintaining learning capacity.
- min_samples_split = 5: Improves robustness by requiring more samples to split a node.
- min_samples_leaf = 2: Ensures adequate samples per leaf node.
- max_features = 'sqrt': Reduces correlation between trees.
- class_weight = 'balanced': Addresses class imbalance automatically.
- random_state = 42, n_jobs = -1: Ensure reproducibility and maximize computational efficiency.

### 3.5 Evaluation Metrics and Validation Strategy

The primary evaluation metric was AUC (Area Under the ROC Curve), which measures discriminative ability independent of classification thresholds, making it well-suited for imbalanced datasets.

To obtain reliable performance estimates, 5-Fold Cross-Validation was applied. The dataset was split into five folds, with each fold serving once as the validation set and the remaining four as training sets. The final reported metrics are the averages across all five folds, ensuring robust and unbiased evaluation.

### 3.6 Multi-Stage Error Diagnosis

The core of this study is a multi-stage diagnosis designed to improve overall error detection accuracy by sequentially filtering different error types, the research process is shown in Figure 1.

- Stage 1: Detects for out_of_range_error. If the model identifies an out_of_range_error, the diagnosis is complete.
- Stage 2: If Stage 1 does not detect an out_of_range error, it then checks for an address_error. If the model identifies an address_error, the diagnosis is complete.
- Stage 3: If Stage 2 does not detect an address_error, it checks for a data_error_0. If the model identifies a data_error_0, the diagnosis is complete.
- Stage 4: If Stage 3 does not detect a data_error_0, it checks for a data_error_1. If the model identifies a data_error_1, the diagnosis is complete.
- Stage 5: If none of the above stages detect a specific error, the final determination is no_error.

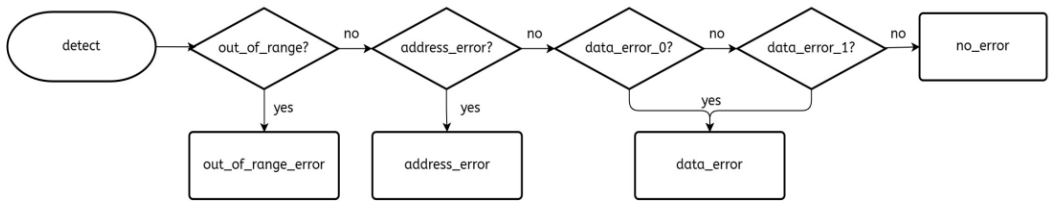

Figure 1: Multi-stage Diagnosis Diagram

## 4 EXPERIMENTAL RESULT

We independently tested the four pre-trained binary classification models and evaluated their performance on their respective error detection tasks. The following are the confusion matrix analyses for each model on the test set.

### 4.1 out_of_range Detection Model

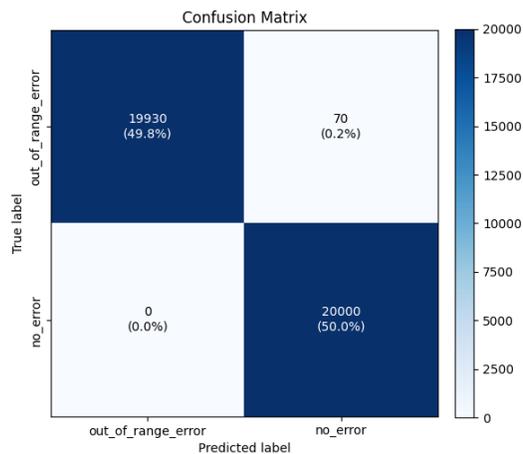

Figure 2: out_of_range Confusion Matrix

As shown in the Figure2 above, the out_of_range detection model performed excellently on the test set, achieving an accuracy of 99.8%.

- True Positive (predicted as out_of_range_error, actual as out_of_range_error): 19,930 instances, accounting for 49.8% of the total.
- True Negative (predicted as no_error, actual as no_error): 20,000 instances, accounting for 50.0% of the total.
- False Positive (predicted as out_of_range_error, actual as no_error): 0 instances, accounting for 0.0% of the total.
- False Negative (predicted as no_error, actual as out_of_range_error): 70 instances, accounting for 0.2% of the total.

In addition, this model achieved an AUC of 1.0000, demonstrating perfect discriminative capability, and a 5-Fold Cross-Validation average accuracy of 0.9940 ± 0.0034, indicating strong generalization ability across different data splits.

This indicates that the model can very accurately identify out_of_range errors while successfully avoiding the misclassification of no_error samples, demonstrating high specificity.

## 4.2 address_error Detection Model

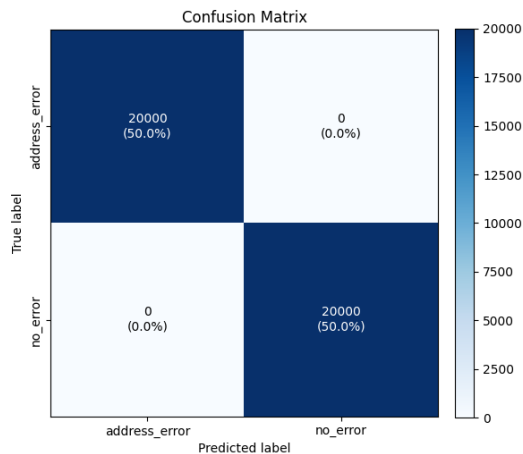

Figure 3: address_error Confusion Matrix

As shown in the Figure 3 above, the performance of the address_error detection model on the test set was ideal, achieving an accuracy of 100%.

- True Positive (predicted as address_error, actual as address_error): 20,000 instances, accounting for 50.0% of the total.
- True Negative (predicted as no_error, actual as no_error): 20,000 instances, accounting for 50.0% of the total.
- False Positive (predicted as address_error, actual as no_error): 0 instances, accounting for 0.0% of the total.
- False Negative (predicted as no_error, actual as address_error): 0 instances, accounting for 0.0% of the total.

The model achieved an AUC of 1.0000 and a 5-Fold Cross-Validation average accuracy of 1.0000 ± 0.0000, showing perfect classification performance and complete consistency across different validation folds. This provides a solid foundation for the early stages of the multi-stage diagnosis.

This model achieved perfect accuracy in identifying address errors, providing a solid foundation for the early stages of the multi-stage diagnosis.

### 4.3 data_error_0 Detection Model

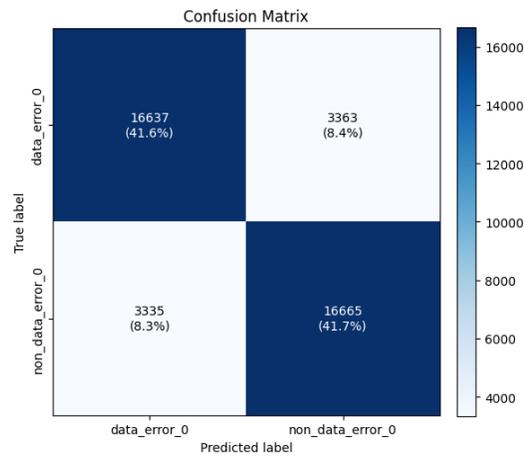

Figure 4: data_error_0 Confusion Matrix

As shown in the figure above, the data_error_0 detection model achieved an accuracy of 83.3% in identifying floating data errors (bits are 00)

- True Positive (predicted as data_error_0, actual as data_error_0): 16,637 instances, accounting for 41.6% of the total.
- True Negative (predicted as non_data_error_0, actual as non_data_error_0): 16,665 instances, accounting for 41.7% of the total.
- False Positive (predicted as data_error_0, actual as non_data_error_0): 3,335 instances, accounting for 8.3% of the total.
- False Negative (predicted as non_data_error_0, actual as data_error_0): 3,363 instances, accounting for 8.4% of the total.

The model obtained an AUC of 0.9095 and a 5-Fold Cross-Validation average accuracy of 0.8177 ± 0.0025, reflecting strong and stable detection capability for specific data error patterns in diverse validation scenarios.

### 4.4 data_error_1 Detection Model

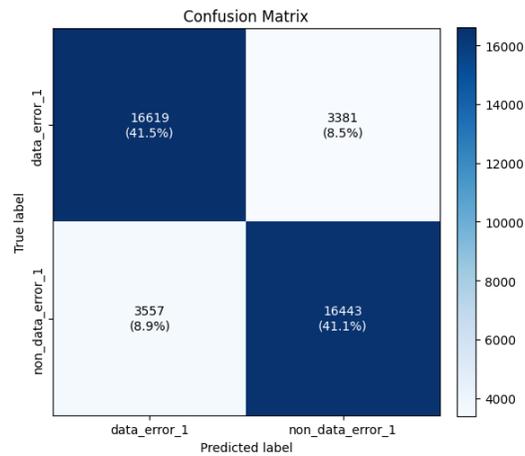

Figure 5: data_error_1 Confusion Matrix

As shown in the figure above, the data_error_1 detection model achieved an accuracy of 82.6% in identifying floating data errors (bits are 11).

- True Positive (predicted as data_error_1, actual as data_error_1): 16,619 instances, accounting for 41.5% of the total.
- True Negative (predicted as non_data_error_1, actual as non_data_error_1): 16,443 instances, accounting for 41.1% of the total.
- False Positive (predicted as data_error_1, actual as non_data_error_1): 3,557 instances, accounting for 8.9% of the total.
- False Negative (predicted as non_data_error_1, actual as data_error_1): 3,381 instances, accounting for 8.5% of the total.

The model achieved an AUC of 0.9039 and a 5-Fold Cross-Validation average accuracy of 0.8156 ± 0.0029, demonstrating reliable performance and consistent recognition capability for targeted data error detection tasks.

## 4.5 Consolidated Results of the Multi-Stage Diagnosis

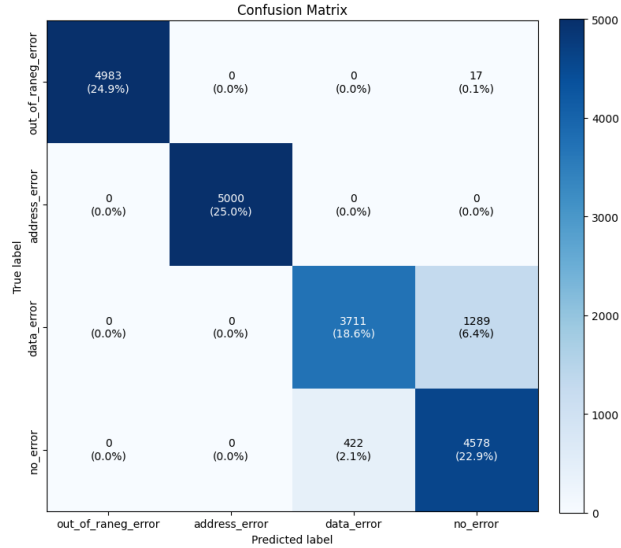

Figure 6: Confusion Matrix of Multi-stage Diagnosis Results

Table 1: Table of Multi-stage Diagnosis Results

| Class | Precision(%) | Recall(%) | F1(%) | TP |
|---|---|---|---|---|
| out_of_range_error | 100.00 | 99.66 | 99.83 | 4,983 |
| address_error | 100.00 | 100.00 | 100.00 | 5,000 |
| data_error | 89.79 | 74.22 | 81.27 | 3,711 |
| no_error | 77.80 | 91.56 | 84.12 | 4,578 |
| Overall accuracy | 91.36% (18,272 / 20,000) | | | |

The consolidated results of the multi-stage diagnosis, as presented in the confusion matrix and performance table (Figures 6 and Table 1), demonstrate its robust capability for end-to-end error detection. The overall accuracy of the method reached 91.36% on the test set, successfully classifying 18,272 out of 20,000 samples. This high overall accuracy underscores the effectiveness of our hierarchical approach.

Analysis of the results for each error class reveals key strengths of the proposed method:

- out_of_range_error: The model achieved a perfect Precision of 100%, indicating that all samples predicted as out_of_range_error were, in fact, correct. With a Recall of 99.66%, it successfully identified 4,983 of the 4,993 true out_of_range_error. This class shows an excellent balance of precision and recall, reflected in an F1 score of 99.83%. This high performance is a significant advantage, as it ensures that this easily identifiable error type is captured with very high confidence.
- address_error: The diagnosis of this class demonstrated flawless performance, achieving 100% Precision, 100% Recall, and a 100% F1 score. This indicates that the system perfectly identified all

5,000 address_error instances. This perfect performance validates the design choice of placing address-type errors early in the diagnosis process, providing a highly reliable initial filter.
- data_error: The diagnosis of data_error types performed effectively, which are inherently more complex to detect. It achieved a high Precision of 89.79%, demonstrating that when the system identifies a data_error, the prediction is correct with a high degree of confidence. The Recall of 74.22% shows the system's strong capability to capture a significant portion of these challenging errors. The solid F1 score of 81.27% confirms that the system is a strong and reliable tool for identifying this complex error class.
- no_error: The diagnosis also effectively classified healthy transactions, achieving a high Recall of 91.56% for the no_error class. This suggests the system is highly effective at not flagging healthy transactions as errors. The Precision for no_error was 77.80%, reflecting the method's ability to maintain a strong distinction between no-error and various error types.

In summary, these results highlight the effectiveness of the multi-stage diagnosis, particularly its exceptional performance on address-related errors and its robust, reliable performance on the more challenging data-related errors. This architecture successfully integrates the strengths of individual models to achieve a high overall accuracy, making it a robust and valuable solution for APB transaction error diagnosis.

## 5 CONCLUSION

The multi-stage diagnosis approach proposed in this study demonstrates superior overall performance and significant practical value on simulated APB transactions. By adopting sequential decision-making architecture, placing address-type errors earlier in the diagnosis process, the system can effectively and robustly capture address-type anomalies, providing a fast and highly confident initial screening for hardware debugging. For data-type errors, the two experimental sub-models (data_error_0 and data_error_1) designed in this experiment, when evaluated together, prove capable of identifying data anomalies with high precision. Although the final result of the ICCAD 2025 CAD Contest is yet to be announced as of the submission date, our team achieved first place in the beta stage, underscoring the method's competitive strength. The overall system achieves an end-to-end recognition rate of 91.36%, fully demonstrating the practical value and scientific merit of this method in the fields of Electronic Design Automation and hardware debugging.

Overall, this experiment validates that a multi-stage, task-oriented sequence of binary classifiers for hierarchical detection of transaction errors is a correct and fruitful technical path. This method exhibits a clear advantage in layering detection difficulty (easier-to-identify address-type errors first, more difficult data-type errors later), providing a robust and compelling starting point for building automated error detection in the EDA toolchain, with broad application prospects.

# 6  LIMITATIONS AND FUTURE WORK

While the proposed hierarchical Random Forest-based diagnostic framework achieved high accuracy on the 2025 ICCAD Contest Problem D benchmark, several limitations remain, which point towards meaningful directions for future advancement.

I. ***Data Validation and Generalization:*** The model was trained on a synthetic dataset created to meet contest rules, not real hardware signals. Its performance on real-world IC design flows is unconfirmed, as such test data is difficult to obtain. To address this, future work should validate the method using real hardware test data to prove its practical utility.

II. ***Limited Scope of Error Types:*** The study was limited to detecting three specific error types defined by the contest (Out-of-Range Access, Address Corruption, and Data Corruption). Its ability to handle other potential protocol errors is unverified. This focused approach was chosen to validate the feasibility of the hierarchical architecture. The framework is flexible and can be expanded to include new classifiers for additional error types as they are defined.

III. ***Feature Engineering Simplicity:*** The model uses Xaddress and Xdata as input features, neglecting more complex temporal and interactive signal features from VCD files. This may limit its ability to capture complex errors. The initial focus was on validating the multi-stage architecture with simple features, and future research can enhance the model's adaptability by incorporating richer feature dimensions and signal interaction characteristics.

IV. ***Cascading Error Propagation:*** The hierarchical structure can propagate misclassifications from early stages, making it impossible for later stages to correct the error. The cascade approach was chosen for its efficiency in filtering simple cases. Future work could introduce multi-path analysis or back-end verification to mitigate the negative effects of early-stage errors.

V. ***Assumed Balanced Data Distribution:*** The model was trained on a balanced dataset with a 50%/50% split of positive and negative samples, which doesn't reflect the highly imbalanced nature of real-world scenarios where errors are rare. While a balanced dataset is useful for initial validation, practical applications would require imbalanced learning strategies, such as data re-weighting, to reduce the discrepancy between the training environment and real-world conditions.

VI. ***Limited Benchmarking:*** The paper focuses on the model's own performance validation and doesn't include a systematic comparison against other methods like rule-based approaches, traditional machine learning models (e.g., SVM), or commercial EDA tools. Due to contest constraints, the study focused on core feasibility. A comprehensive comparison with similar open-source algorithms and commercial tools is necessary to fully establish the method's competitiveness and industrial value.

# Authors' background

| Your Name | Title* | Research Field | Personal website |
|---|---|---|---|
| Cheng-Yang Tsai | Undergraduate Student | EDA, Algorithm | N/A |
| Tzu-Wei Huang | Undergraduate Student | EDA, Algorithm | N/A |
| JEN-WEI SHIH | Undergraduate Student | EDA, Algorithm | N/A |
| I-HSIANG WANG | Undergraduate Student | EDA, Algorithm | N/A |
| Yu-Cheng Lin | Assistant Professor | EDA, Algorithm, AI music | https://www.cse.yzu.edu.tw/en/people/professor?name=Yu-Cheng%20Lin |
| RUNG-BIN LIN | Professor | VLSI, EDA, IC design, Computer architecture | https://www.cse.yzu.edu.tw/en/people/professor?name=Rung-Bin%20Lin |